\documentclass[aps,prl,twocolumn,superscriptaddress,longbibliography,floatfix]{revtex4-2}
\usepackage{graphicx,bm,amssymb,amsmath,dcolumn,hyperref}
\usepackage{multirow}
\usepackage{enumerate}
\usepackage{enumitem}
\usepackage{color}
\usepackage{subfigure}  
\usepackage{epstopdf}
\usepackage{bbm}
\usepackage{graphicx}
\usepackage{threeparttable}
\usepackage{amsthm}
\usepackage{mathtools}
\usepackage{color}
\usepackage{tikz}
\usepackage{float}
\usepackage{empheq}     
\usepackage{booktabs}
\usepackage{natbib}

\draft
\graphicspath{{./images/}}

\begin{document}

\preprint{APS/123-QED}

\title{Criticality and Universality of Generalized Kuramoto Model}

\author{Zhongpu Qiu}
\affiliation{School of Systems Science/Institute of Nonequilibrium Systems, Beijing Normal University, Beijing 100875, China}%

\author{Tianyi Wu}
\affiliation{School of Systems Science/Institute of Nonequilibrium Systems, Beijing Normal University, Beijing 100875, China}%

\author{Sheng Fang}
\affiliation{School of Systems Science/Institute of Nonequilibrium Systems, Beijing Normal University, Beijing 100875, China}%

\author{Jun Meng}
\affiliation{Key Laboratory of Earth System Numerical Modeling and Application, Institute of Atmospheric Physics, Chinese Academy of Sciences, Beijing 100029, China}

\author{Jingfang Fan}
\email{jingfang@bnu.edu.cn}
\affiliation{School of Systems Science/Institute of Nonequilibrium Systems, Beijing Normal University, Beijing 100875, China}
\affiliation{Potsdam Institute for Climate Impact Research, Potsdam 14412, Germany}

\begin{abstract}
We explore synchronization transitions in even-$D$-dimensional generalized Kuramoto oscillators on both complete graphs and $d$-dimensional lattices. In the globally coupled system, analytical expansions of the self-consistency equations, incorporating finite-size corrections, reveal universal critical exponents $\beta = 1/2$ and $\bar{\nu} = 5/2$ for all even $D$, indicating an unconventional upper critical dimension $d_u = 5$. Extensive numerical simulations across multiple $D$ confirm these theoretical predictions. For locally coupled systems, we develop a framework based on spin-wave theory and fluctuation-resolved functional network diagnostics, which captures criticality in entrainment transition.  A modified Edwards-Anderson order parameter further validates the predicted exponents. This combined theoretical and numerical study uncovers a family of universality classes characterized by $D$-independent but $d$-dependent criticality, offering a unified perspective on symmetry and dimensionality in nonequilibrium synchronization phenomena.
\end{abstract}

\maketitle

\paragraph{Introduction}
Synchronization is a universal collective phenomenon observed across a wide range of natural and engineered systems, from flashing fireflies~\cite{sarfati_self-organization_2021} and chemical oscillators~\cite{Kiss_2002} to superconducting Josephson junctions~\cite{wiesenfeld_synchronization_1996} and beyond~\cite{pikovsky_synchronization:_2003}. The Kuramoto model~\cite{kuramoto_self-entrainment_1975,kuramoto_chemical_1984,strogatz_kuramoto_2000,boccaletti_synchronization_2018}, a paradigm of globally coupled phase oscillators with distributed natural frequencies, whose analytical accessibility has been central to elucidating synchronization transitions,  drawing deep analogies with critical phenomena in statistical mechanics~\cite{nishimori_elements_2010,dorogovtsev_critical_2008}.

Beyond the mean-field limit, spatially extended systems exhibit richer and often nontrivial collective behaviors. Synchronization may emerge abruptly through the formation of macroscopic clusters with statistically identical frequencies—a phenomenon known as the entrainment transition~\cite{acebron_kuramoto_2005}. Its nature depends sensitively on interaction topology, as demonstrated across complete graphs~\cite{hong_entrainment_2007,hong_finite-size_2015}, regular lattices~\cite{sakaguchi_local_1987,strogatz_phase-locking_1988,daido_lower_1988,daido_susceptibility_2015}, and complex networks~\cite{rodrigues_kuramoto_2016,arenas_synchronization_2008,arenas_synchronization_2006,del_genio_synchronization_2016}. These studies reveal that spatial dimensionality and interaction topology play a central role in shaping the universality class of the synchronization transition, while the underlying phase oscillators are fixed to the classical Kuramoto dynamics~\cite{tang_synchronize_2011}.

Recent developments have extended the classical Kuramoto model to $D$-dimensional vectorial oscillators~\cite{olfati-saber_swarms_2006,Zhu_2013,Tanaka_2014}, leading to the so-called $D$-dimensional Kuramoto model (DDKM). This generalization exhibits a striking parity-induced dichotomy in critical behavior: even-$D$ systems undergo continuous transitions, while odd-$D$ counterparts display discontinuities~\cite{chandra_continuous_2019,chandra_complexity_2019}. Further studies have investigated the effects of positive feedback~\cite{dai_discontinuous_2020}, amplitude dynamics~\cite{zou_solvable_2023,wang_rhythmic_2024}, matrix-valued couplings~\cite{buzanello_matrix_2022}, and network generalizations including complex and higher-order topologies~\cite{ling_effects_2022,dai_d-dimensional_2021,kovalenko_contrarians_2021,wang_higher-order_2025}.

Despite these advances, a fundamental question remains unresolved: How do symmetry (determined by the order parameter dimension $D$) and spatial dimensionality $d$ jointly govern the universality class of synchronization transitions? Universality is a cornerstone of modern statistical physics, enabling the classification of systems that, despite microscopic diversity, exhibit identical scaling behavior near critical points~\cite{stanley_scaling_1999}. In equilibrium systems, such classifications have established unifying principles across magnetism and thermal phase transitions, grounded in symmetry and dimensionality. Although the Kuramoto model exhibits formal analogies with these transitions, it fundamentally represents a nonequilibrium system composed of dissipative self-driven oscillators that do not satisfy detailed balance. Compared to equilibrium theory, the universality structure of nonequilibrium systems remains poorly understood, especially in contexts involving vectorial synchronization in coupled $D$-dimensional oscillators. Clarifying the $(D,d)$-dependence of critical exponents and identifying novel universality classes is therefore essential for constructing a coherent theoretical framework for collective dynamics far from equilibrium.

In this Letter, we address how symmetry and spatial dimensionality jointly determine the universality class of synchronization transitions by studying coupled even-$D$-dimensional phase oscillators in both globally and locally coupled systems with finite spatial dimension $d$. For globally coupled DDKM, we analytically expand the finite-size self-consistency equation and confirm the predictions through extensive numerical simulations. We uncover a universal set of critical exponents for all even $D$, including a correlation size exponent $\bar{\nu} = 5/2$, implying an unconventional upper critical dimension $d_u = 5$. To investigate the impact of spatial dimensionality (locally coupled DDKM), we apply spin-wave theory and introduce the \textit{entrainment connectivity matrix}, which represents a functional network constructed from long-time fluctuations in pairwise synchronization errors. A modified Edwards-Anderson order parameter is developed to validate critical exponents. This framework reveals the $(D,d)$-dependence of both critical coupling strengths and exponents in even-$D$ systems, as summarized in Table.~\ref{tab:Criticality_table_1}.

\paragraph{Globally coupled oscillators.}
We begin with the DDKM under global coupling, described by~\cite{chandra_continuous_2019},
\begin{equation}
\frac{d\vec{\sigma}_i}{dt} = K\left[\vec{\rho} - (\vec{\sigma}_i \cdot \vec{\rho})\vec{\sigma}_i\right] + \mathbf{W}_i \vec{\sigma}_i,
\label{eq:Criticality_1}
\end{equation}
where $\vec{\sigma}_i \in \mathbb{R}^D$ is the unit state vector of the $i$-th oscillator, $K$ is the coupling strength, and $\mathbf{W}_i$ is a random $D \times D$ antisymmetric matrix with independent standard Gaussian upper triangular elements. The collective dynamics are characterized by the phase order parameter $\vec{\rho} = N^{-1} \sum_i \vec{\sigma}_i$, with $\rho = |\vec{\rho}|$ quantifying synchronization.

Analysis with antisymmetric matrix block diagonalization of the fixed points of Eq.~(\ref{eq:Criticality_1}) reveals the phase-locking condition for each oscillator~\cite{kuramoto_self-entrainment_1975,chandra_continuous_2019,Criticality_SM},
\begin{equation}
K^2 \sum_{k=1}^{D/2} \frac{\rho_k^2}{w_{i,k}^2} > 1,
\label{eq:Criticality_2}
\end{equation}
where $\rho_k^2$ is the sum of the squares of the $(2k-1)$-th and $2k$-th components of the basis-transformed order parameter and $w_{i,k}$ is the $k$-th eigenvalue of $\mathbf{W}_i$. When this condition is satisfied, the oscillator phase-locks and contributes coherently to $\rho$; otherwise, it drifts incoherently and contributes negligibly on average.

For finite $N$, sample-to-sample fluctuations introduce corrections to the self-consistency equation of $\rho$, which can be expressed as~\cite{hong_finite-size_2015,tang_synchronize_2011},
\begin{equation}
\rho^2 = \langle \lambda \rangle + \sqrt{\frac{\langle \lambda^2 \rangle - \langle \lambda \rangle^2}{N}} \eta,
\label{eq:Criticality_3}
\end{equation}
where $\eta$ is a standard Gaussian random variable and $\lambda=\rho_1 \sqrt{1 - (\omega_1/K \rho_1)^2}$ represents the contribution to $\rho^2$ of individual oscillator in one of effective phase-locked subregions by simplifing Eq.~(\ref{eq:Criticality_2})~\cite{Criticality_SM}. Here $\langle\cdot\rangle$ denotes ensemble average. The mean $\langle\lambda\rangle = \tilde{g}_0(0) C_0 K \rho^2 + 4\tilde{g}_2(0) C_1 K^3 \rho^4/(D+2)$ and the variance $\langle\lambda^2\rangle - \langle\lambda\rangle^2 = \tilde{g}_0(0) D_0 K \rho^3$ arise from integrating over reduced phase-locked regions using the Beta function $B(x,y)$, with $C_m = B(m+\frac{1}{2},\frac{3}{2})$, $D_m = B(m+\frac{1}{2},2)$ and $\tilde{g}_n(0)$ represents integral constant~\cite{Criticality_SM}.

Near $K_c$, Eq.~(\ref{eq:Criticality_3}) simplifies to a nonlinear stochastic equation for $\rho$~\cite{Criticality_SM} with constants $a$ and $b$:
\begin{equation}
\frac{K - K_c}{K_c} \rho + a K^3 \rho^3 + b \left(\frac{K}{N}\right)^{1/2} \rho^{1/2} \eta = 0,
\label{eq:Criticality_4}
\end{equation}
where $K_c = 1/(\tilde{g}_0(0)C_0)$ denotes the critical coupling.

Applying finite-size scaling $\rho = N^{-\beta/\bar{\nu}} f((K-K_c)N^{1/\bar{\nu}})$~\cite{privman1990finite} and comparing exponents yields the relations~\cite{Criticality_SM},
\begin{equation}
-\frac{\beta}{\bar{\nu}} - \frac{1}{\bar{\nu}} = -\frac{3\beta}{\bar{\nu}} = -\frac{\beta}{2\bar{\nu}} - \frac{1}{2},
\label{eq:Criticality_5}
\end{equation}
which uniquely determine the critical exponents,
\begin{equation}
\beta = \frac{1}{2}, \quad \bar{\nu} = \frac{5}{2}.
\label{eq:Criticality_6}
\end{equation}

These two exponents govern the scaling of the order parameter and the correlation size, respectively. Notably, the anomalous value $\bar{\nu} = 5/2$ deviates markedly from the equilibrium mean-field value $\bar{\nu} = 2$~\cite{nishimori_elements_2010}, reflecting strong sample-to-sample fluctuations inherent in finite systems~\cite{hong_finite-size_2015,tang_synchronize_2011}. Using the relation $\bar{\nu} = \nu_{\text{MF}} d_u = d_u / 2$~\cite{botet_size_1982}, this result implies an unconventional upper critical dimension $d_u = 5$ for all even-DDKM .

\begin{figure}[htbp]
  \centering
  \includegraphics[width=0.5\textwidth]{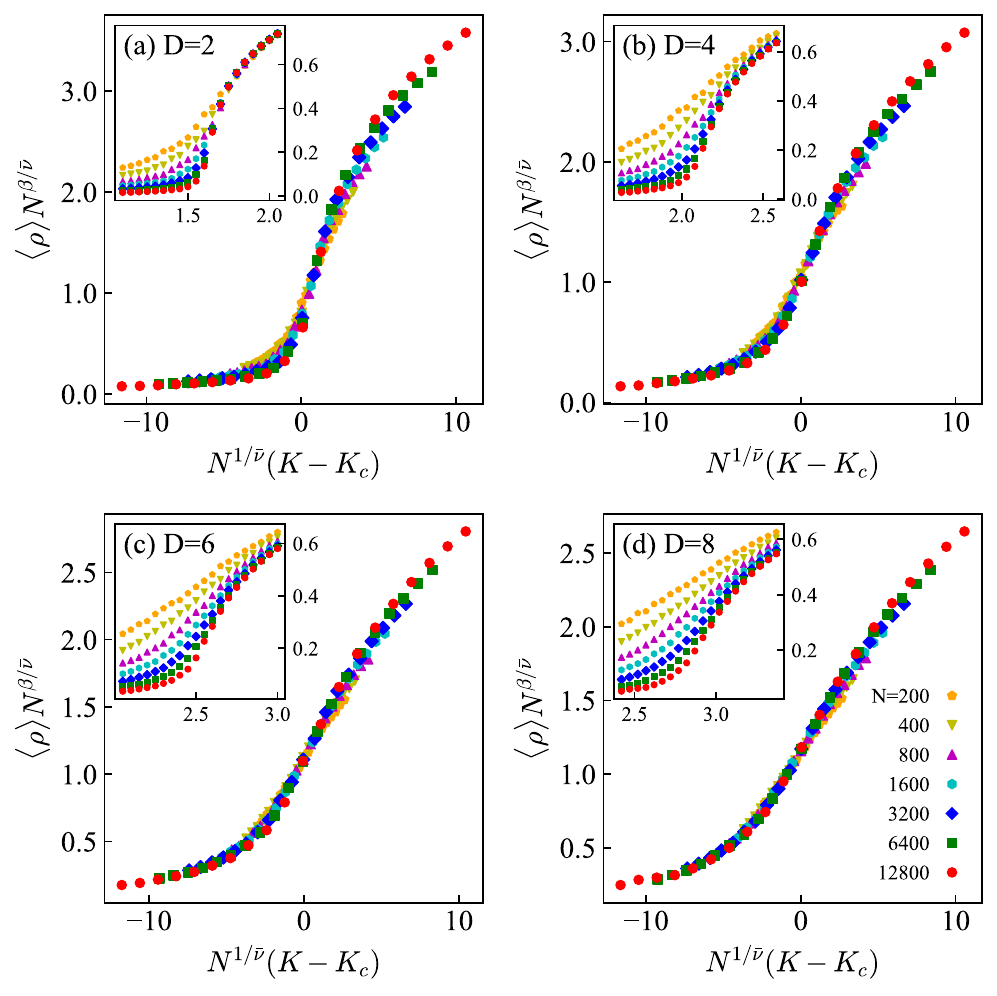}
  \caption{Finite-size scaling analysis of the order parameter $\rho$ in even-DDKM for $D = 2, 4, 6, 8$ [(a)–(d)]. Insets: $\rho$ as a function of $K$ for various system sizes $N$. Main panels: scaled plots confirm universal critical exponents $\beta = 1/2$ and $\bar{\nu} = 5/2$, demonstrating $D$-independent criticality across all even-$D$ cases.}
  \label{fig:figure_1}
\end{figure}

These findings generalize earlier results on the classical ($D=2$) Kuramoto model~\cite{hong_entrainment_2007}, demonstrating that all even-$D$ globally coupled models fall within the same universality class. Extensive numerical simulations confirm the predicted scaling behavior, as shown in Fig.~\ref{fig:figure_1}.

\begin{figure*}[htbp]
  \centering
  \includegraphics[width=0.95\textwidth]{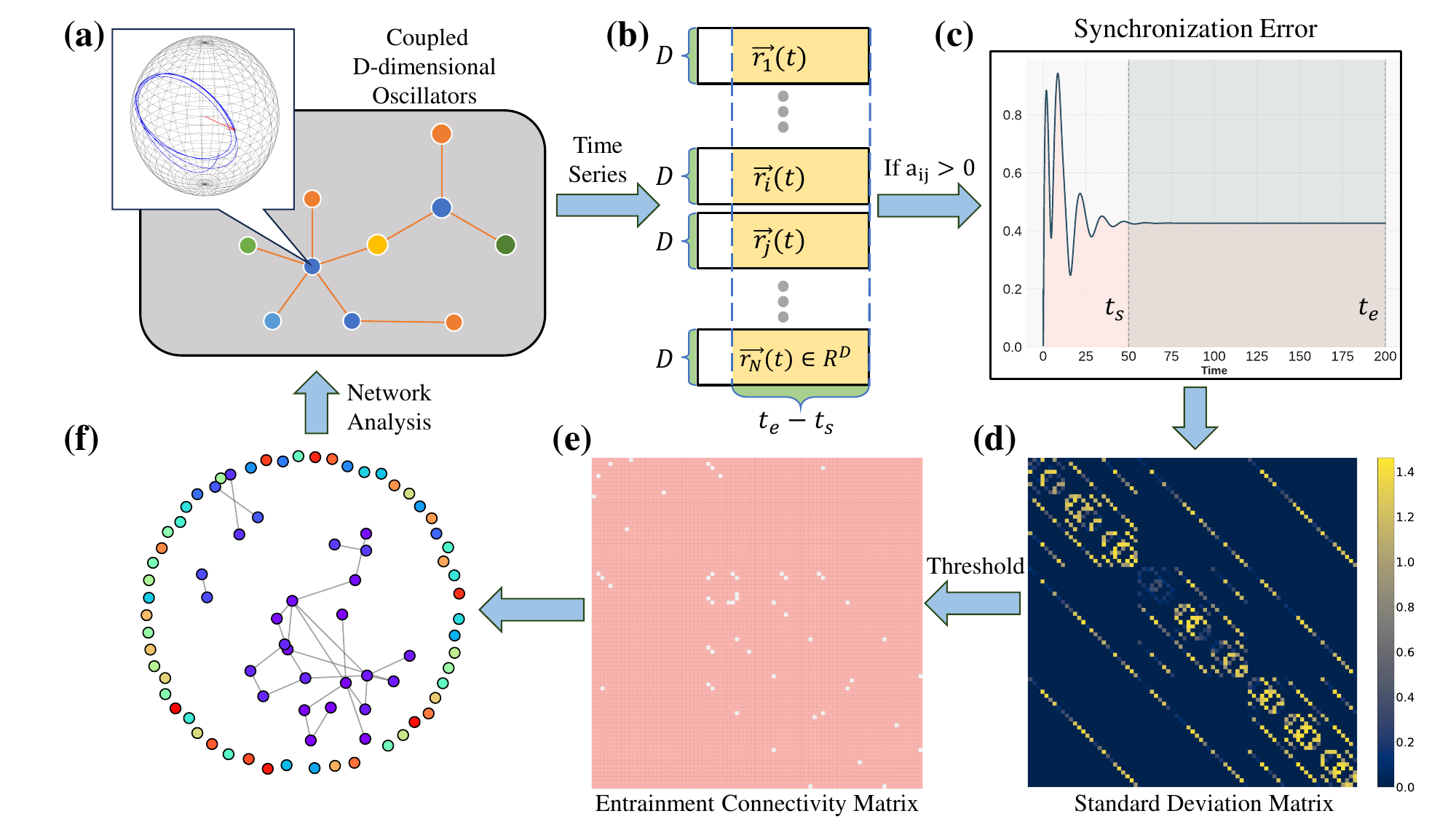}%
  \hfill
  \parbox[b]{1\textwidth}{%
    \caption{Illustration of the framework for identifying entrained clusters in generalized DDKM dynamics. (a) Network of $D$-dimensional coupled oscillators. (b) Time series of state vectors are extracted for each node. (c) Synchronization error $e_{ij}(t)$ is computed for connected pairs $(i,j)$ over a long interval $[t_s, t_e]$ to exclude transients and ensure statistical accuracy. (d) The standard deviation $\sigma_{ij}$ of $e_{ij}(t)$ quantifies dynamical coherence. (e) An entrainment connectivity matrix (ECM) is constructed by thresholding $\sigma_{ij}$ at $\epsilon = 0.1$; ECM entries are 1 if $\sigma_{ij} < \epsilon$ and 0 otherwise. (f) The ECM is analyzed topologically to extract entrained clusters and other network characteristics relevant to synchronization dynamics. For example, the number of zero eigenvalues of the ECM corresponds to the number of entrained clusters in the system.}

    \label{fig:example}%
  }
\end{figure*}

\begin{figure}[htbp]
  \centering
  \includegraphics[width=0.48\textwidth]{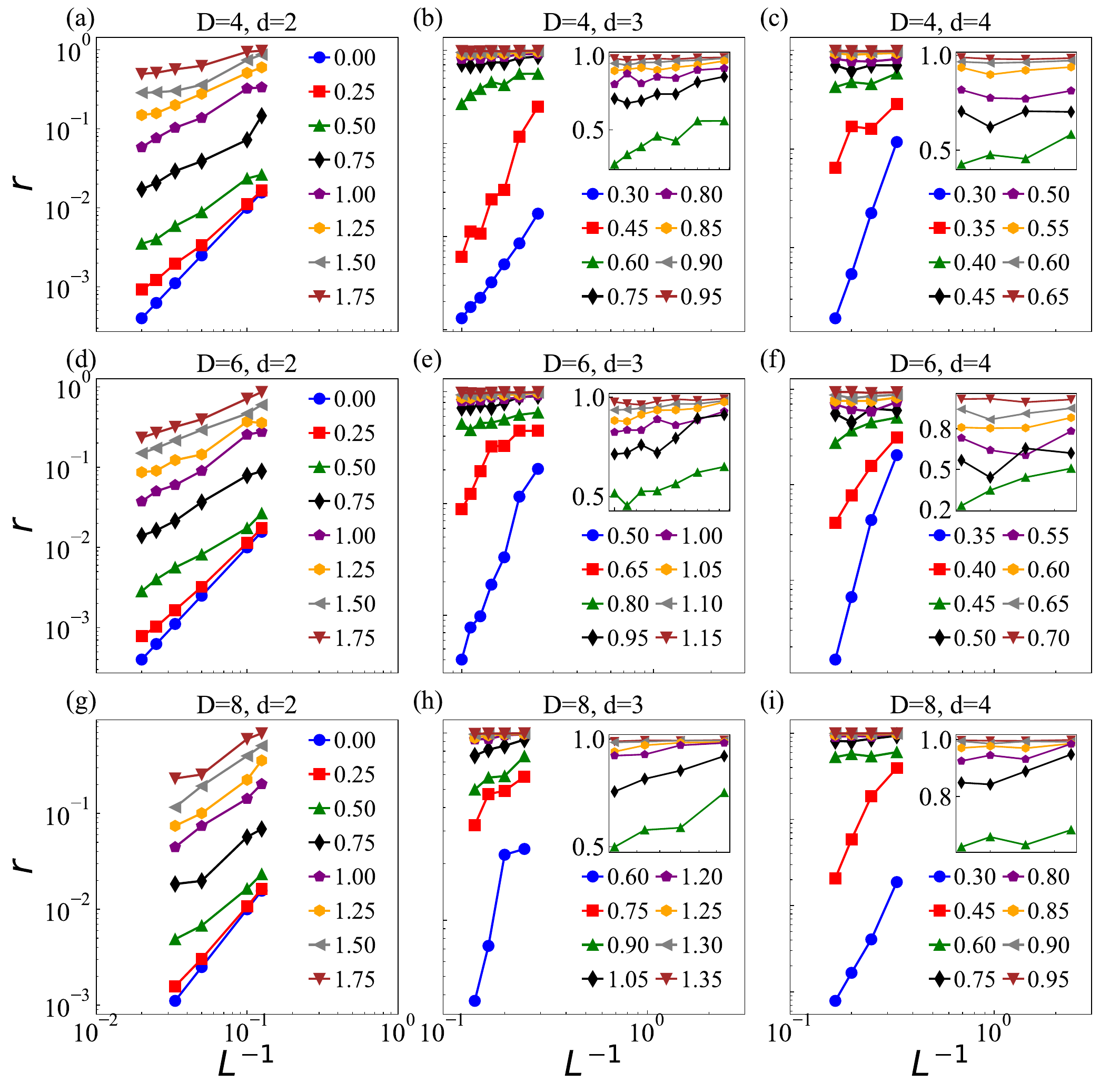}
 \caption{Dependence of the frequency order parameter $r$ on inverse system size $L^{-1}$ for various coupling strengths $K$ in systems with $D = 4, 6, 8$ and spatial dimensions $d = 2, 3, 4$ [(a)–(i)]. Insets: magnified views highlighting critical behavior in the $d = 3$ and $d = 4$ cases.}
  \label{fig:figure_3}
\end{figure}

\paragraph{Locally coupled oscillators.}
We now consider DDKM dynamics on a $d$-dimensional hypercubic lattice, where each oscillator interacts with its nearest neighbors. In this locally coupled setting, Eq.~(\ref{eq:Criticality_1}) becomes,
\begin{equation}
\frac{d\vec{\sigma}_i}{dt} =  K\sum_{j \in \Lambda_i} \left[\vec{\sigma}_j - (\vec{\sigma}_j \cdot \vec{\sigma}_i) \vec{\sigma}_i\right] + \mathbf{W}_i \vec{\sigma}_i,
\label{eq:Criticality_7}
\end{equation}
where $\Lambda_i$ denotes the nearest-neighbor set of site $i$.

To quantify local synchronization, we define the pairwise synchronization error (SE) between oscillators $i$ and $j$ as,
\begin{equation}
e_{ij}(t) = \|\vec{\sigma}_i(t) - \vec{\sigma}_j(t)\|^2,
\label{eq:Criticality_8}
\end{equation}
which generalizes the phase difference used in classical ($D = 2$) Kuramoto models~\cite{dsouza_explosive_2019,ghosh_amplitude_2024,ravoori_robustness_2011,zhang_critical_2020,zhou_hierarchical_2006}.


In the strong-coupling limit ($K \to \infty$), we employ the spin-wave approximation by linearizing fluctuations around the entrained state. This approach yields analytical predictions for the spatial scaling of ensemble average of the steady-state synchronization error $E_{ij} = \langle e_{ij}(t) _{t \to \infty}\rangle$ as a function of oscillator separation $\mathbf{r}_{ij}$ and system size $L$~\cite{Criticality_SM}:
\begin{equation}
E_{ij} \sim
\begin{cases}
|\mathbf{r}_{ij}|^2 L^{-(d-2)}, & d < 2 \\[0.5em]
|\mathbf{r}_{ij}|^2 \ln(L/|\mathbf{r}_{ij}|), & d = 2 \\[0.5em]
|\mathbf{r}_{ij}|^{4-d}, & 2 < d < 4 \\[0.5em]
1/2, & d = 4 \\[0.5em]
0. & d > 4
\end{cases}
\label{eq:Criticality_9}
\end{equation}

These scalings reveal three critical regimes\cite{Criticality_SE}:
(i) For $d \leq 2$, long-range synchronization is suppressed due to diverging fluctuations, establishing the lower critical dimension $d_l = 2$. In this regime, the synchronization error grows with system size, indicating the absence of both frequency entrainment and phase coherence in thermodynamic limit.
(ii) For $2 < d \leq 4$, frequency entrainment emerges without phase order. The SE diverges with oscillator separation as $|\mathbf{r}_{ij}|^{4-d}$, while the correlation length $\xi$, which characterizes the typical spatial scale over which oscillator frequencies remain entrained, scales as $\xi \sim \delta K^{-2/(d-2)}$. This yields critical exponents $\nu = 2$ for $d = 3$ and $\nu = 1$ for $d = 4$.
(iii) For $d > 4$, both frequency and phase synchronization are achieved, accompanied by spontaneous symmetry breaking of the $SO(D)$ rotational symmetry for even $D$. These results extend known behaviors of classical Kuramoto oscillators~\cite{sakaguchi_local_1987,hong_collective_2004,hong_collective_2005} to higher-dimensional vectorial dynamics, establishing a $D$-independent spatial dependence in even-$D$ systems.

\begin{figure}[htbp]
  \centering
  \includegraphics[width=0.48\textwidth]{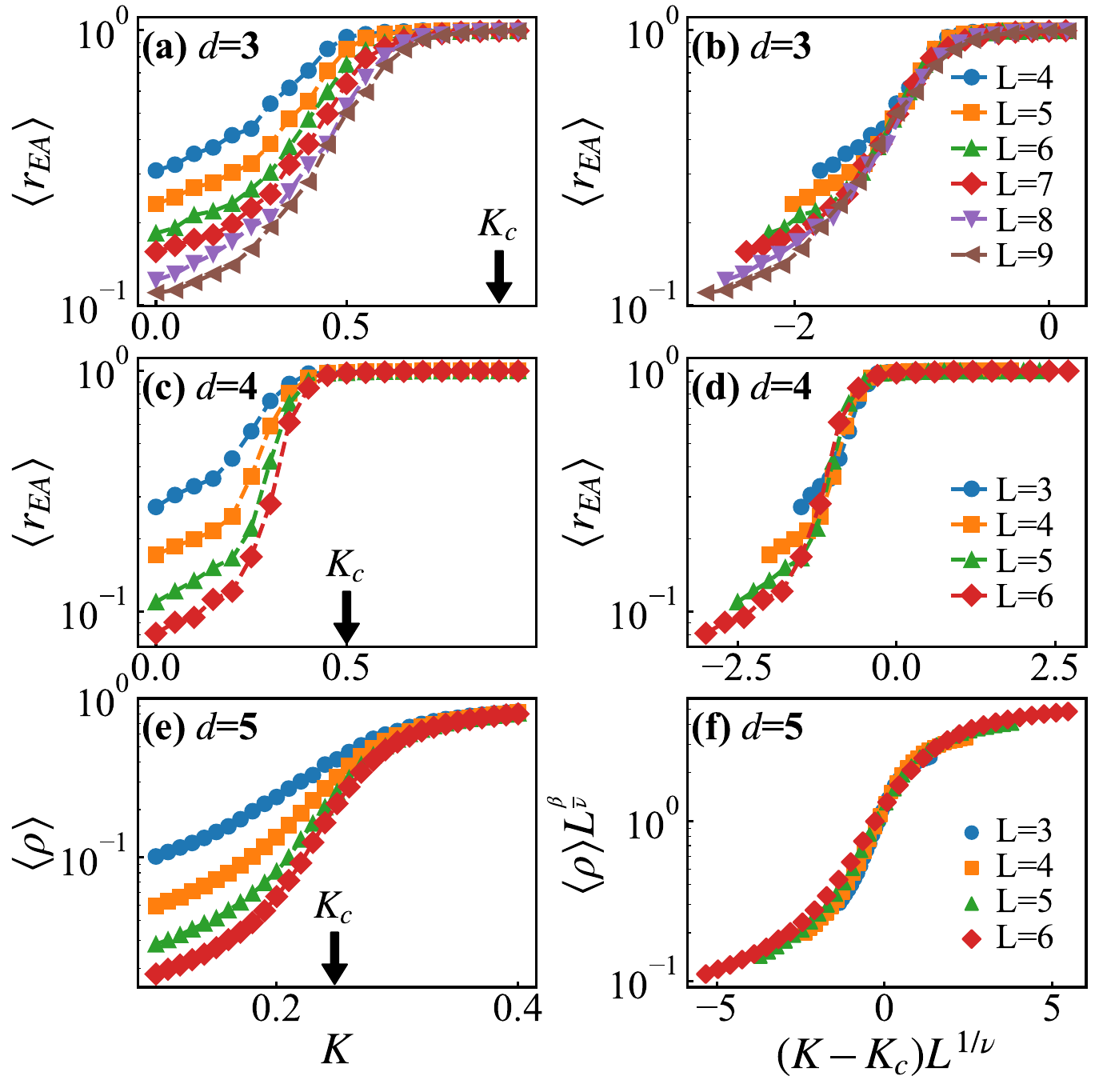}
  \caption{Finite-size scaling of the generalized Edwards-Anderson order parameter $r_{\text{EA}}$ for $D = 4$ oscillators. (a, b) Scaling results for $d = 3$, confirming the predicted critical exponent $\nu = 2$. (c, d) Results for $d = 4$, consistent with $\nu = 1$. (e, f) Results for $d = 5$, exhibiting mean-field scaling of the phase order parameter $\langle \rho \rangle$ with exponents $\beta = 1/2$ and $\nu = 1/2$.}
  \label{fig:figure_4}
\end{figure}

\paragraph{Generalized Entrainment Framework.}
While the linear spin-wave approximation reveals the existence of synchronization transitions, identifying the critical coupling strength in the nonlinear regime for even-$D$ systems requires a robust definition of frequency entrainment beyond $D = 2$. For classical Kuramoto oscillators, entrainment is typically defined via the equality of effective frequencies, $\tilde{\omega} \equiv \lim_{T \to \infty} \frac{1}{T} \int_0^T \dot{\theta}(t)\, dt$~\cite{sakaguchi_local_1987,acebron_kuramoto_2005,mori_necessary_2010,lee_vortices_2010}. However, this definition breaks down for $D$-dimensional oscillators due to the fundamental structure of the special orthogonal group $SO(D)$. While $SO(2)$ admits a single-frequency representation, the dimension of $SO(D)$ grows as
\begin{equation}
\dim SO(D) = \frac{D(D-1)}{2} > D = \dim \mathbb{R}^D,
\end{equation}
creating an underdetermined system for $D>3$. This profound limitation makes direct frequency calculations infeasible.
To overcome this challenge, we propose a framework based on temporal fluctuations of SE, illustrated in Fig.~\ref{fig:example}. Specifically, the standard deviation of the pairwise SE defined as Eq.~(\ref{eq:Criticality_8}),
\begin{equation}
\sigma_{ij} = \sqrt{[(e_{ij}(t) - [e_{ij}(t)])^2]},
\label{eq:Criticality_10}
\end{equation}
(where $[\cdot]$ denotes time averaging) quantifies the stability of dynamical coherence. By applying a threshold to $\sigma_{ij}$~\cite{Criticality_SM} to binarize, we construct an \textit{entrainment connectivity matrix}, representing a functional network of synchrony relationships. The largest connected component in this network corresponds to the largest entrained cluster, from which we compute a percolation-like order parameter $r = N_s/N$ representing the existence of the percolating frequency-entrained cluster, where $N_s$ is the cluster size. In the entrained phase ($K > K_c$), $r$ saturates to a finite value as $L \to \infty$, while in the detrained phase ($K < K_c$), it vanishes. This methodology is validated numerically across complete graphs and lattices with varying $D$ and $d$~\cite{Criticality_SM}, as shown in Fig.~\ref{fig:figure_3}. Inspired by approaches in brain and climate functional networks~\cite{fan_statistical_2021,zhou_hierarchical_2006-1,faskowitz_edge-centric_2020}, this framework provides both a definitional and technical nontrivial generalization of frequency entrainment to $D$-dimensional oscillators. Although similar methods have been used for phase synchronization in $D = 2$ systems~\cite{arenas_synchronization_2006,gomez-gardenes_paths_2007,arenas_synchronization_2008}, our approach is the first to generalize this analysis to entrainment and arbitrary $D$.

To validate the predicted critical exponents for $d = 3$ and $d = 4$, we introduce a modified Edwards-Anderson (EA) order parameter, since the standard form $\Delta_{\text{EA}} = \lim_{T \to \infty} \frac{1}{N} \left| \sum_j e^{i[\phi_j(t_0+T) - \phi_j(t_0)]} \right|$ is only applicable to $D = 2$~\cite{hong_entrainment_2007}. Our modified EA parameter is defined as
\begin{equation}
r_{\text{EA}} = \frac{\left\| \lim_{T \to \infty} \frac{1}{N} \sum_{i=1}^N \vec{\sigma}_i(t_0 + T)\, \vec{\sigma}_i(t_0)^T \right\|_{\text{op}}}{\left\| \frac{1}{N} \sum_{i=1}^N \vec{\sigma}_i(t_0)\, \vec{\sigma}_i(t_0)^T \right\|_{\text{op}}},
\label{eq:Criticality_11}
\end{equation}
where $t_0$ is the initial transient time, and $\| \cdot \|_{\text{op}}$ denotes the spectral norm~\cite{Criticality_EA}. This order parameter effectively captures the long-term directional correlation of oscillators. As shown in Fig.~\ref{fig:figure_4}, $r_{\text{EA}}$ successfully captures the entrainment transition. The extracted critical couplings $K_c$ vary with dimension $D$, but the critical exponents agree with our theoretical predictions, confirming the $D$-independent universality. These results are summarized in Table.~\ref{tab:Criticality_table_1}, demonstrating the consistency and generality of our framework.

\begin{table}[htbp]
\caption{\label{tab:Criticality_table_1}Critical coupling strengths $K_c$ and critical exponents for synchronization transitions across spatial dimensions $d$ and order parameter dimensions $D$. Reported exponents include the order parameter exponent $\beta$, the correlation size exponent $\bar{\nu}$, and the correlation length exponent $\nu$.}
  \begin{ruledtabular}
  \begin{tabular}{cccccc}
  \textrm{$d$} & \textrm{$D = 2$} & \textrm{$D = 4$} & \textrm{$D = 6$} & \textrm{$D = 8$} & \textrm{Quantities}  \\
  \colrule
  $\infty$ & \(\begin{array}{c}
  1/2 ~\cite{kuramoto_chemical_1984} \\
  5/2 ~\cite{hong_entrainment_2007} \\
  1.595 ~\cite{kuramoto_chemical_1984}
  \end{array}\) & \(\begin{array}{c}
  1/2  \\
  5/2  \\
  2.127 ~\cite{chandra_continuous_2019}
  \end{array}\) & \(\begin{array}{c}
  1/2 \\
  5/2 \\
  2.553 ~\cite{chandra_continuous_2019}
  \end{array}\) & \(\begin{array}{c}
  1/2 \\
  5/2 \\
  2.917 ~\cite{chandra_continuous_2019}
  \end{array}\) & \(\begin{array}{c}
  \beta \\
  \bar{\nu} \\
  K_c
  \end{array}\) \\
  \colrule
  
  \multicolumn{6}{c}{...} \\
  \colrule
  \colrule
  
  5        & \(\begin{array}{c}
  1/2 ~\cite{hong_entrainment_2007} \\
  1/2 ~\cite{hong_entrainment_2007} \\
  0.201 ~\cite{hong_entrainment_2007}
  \end{array}\) & \(\begin{array}{c}
  1/2 \\
  1/2 \\
  0.246(2)
  \end{array}\) & \(\begin{array}{c}
  1/2 \\
  1/2 \\
  0.300(2)
  \end{array}\) & \(\begin{array}{c}
  1/2 \\
  1/2 \\
  0.350(2)
  \end{array}\) & \(\begin{array}{c}
  \beta \\
  \nu \\
  K_c
  \end{array}\) \\
  \colrule
  4        & \(\begin{array}{c}
  0 ~\cite{hong_entrainment_2007} \\
  1 ~\cite{hong_entrainment_2007} \\
  0.318 ~\cite{hong_entrainment_2007}
  \end{array}\) & \(\begin{array}{c}
  0 \\
  1 \\
  0.50(5)
  \end{array}\) & \(\begin{array}{c}
  0 \\
  1 \\
  0.60(5)
  \end{array}\) & \(\begin{array}{c}
  0 \\
  1 \\
  0.75(5)
  \end{array}\) & \(\begin{array}{c}
  \beta \\
  \nu \\
  K_c
  \end{array}\) \\
  \colrule
  3        & \(\begin{array}{c}
  0 ~\cite{hong_entrainment_2007} \\
  2 ~\cite{hong_entrainment_2007} \\
  0.80 ~\cite{hong_entrainment_2007}
  \end{array}\) & \(\begin{array}{c}
  0 \\
  2 \\
  0.90(5)
  \end{array}\) & \(\begin{array}{c}
  0 \\
  2 \\
  1.15(5)
  \end{array}\) & \(\begin{array}{c}
  0 \\
  2 \\
  1.35(5)
  \end{array}\) &
  \(\begin{array}{c}
  \beta \\
  \nu \\
  K_c
  \end{array}\) \\
  \colrule
  2        & --- ~\cite{daido_lower_1988} & ---& ---& --- \\
  \end{tabular}
  \end{ruledtabular}
  \end{table}

\paragraph{Conclusion.}
In summary, by systematically exploring the $(D, d)$ parameter space of coupled even-$D$-dimensional phase oscillators, we identify a family of $d$-dependent but $D$-independent universality classes governing the criticality of synchronization transitions. For globally coupled systems, all even-$D$ models exhibit identical critical exponents $(\beta, \bar{\nu}) = (1/2, 5/2)$, indicating a nontrivial upper critical dimension $d_u = 5$ that distinguishes conventional equilibrium spin models. For locally coupled oscillators, our generalized entrainment framework transcends phase-based methods and provides a robust diagnostic for identifying critical behavior across spatial dimensions and network topologies.

This framework is broadly applicable to systems with complex coupling architectures, including random graphs, multilayer networks, and higher-order interactions. Potential applications range from collective motion in biological ensembles~\cite{vicsek_collective_2012,chate_dry_2020,okeeffe_oscillators_2017,yoon_sync_2022} to consensus dynamics in distributed multi-agent systems~\cite{boccaletti_synchronization_2018,peng2023exponential,peng_global_2025}, and even to collective decision-making in social networks. Future work may extend this foundation to study synchronization in structurally complex networks, elucidate the distinct nature of odd-$D$ systems, and analyze dynamical fluctuations and critical responses to external force.

{\section*{Acknowledgements}}
We thank Dr. Sarthak Chandra, Prof. Youjin Deng, Prof. Yueheng Lan, Prof. Xiaqing Shi, Dr. Maoxin Liu and Prof. Jürgen Kurths for valuable discussions and insights. This work was supported by the National Natural Science Foundation of China (Grant No. 42450183, 12275020, 12135003, 12205025, 42461144209) and the Ministry of Science and Technology of China (2023YFE0109000). J.F. is supported by the Fundamental Research Funds for the Central Universities.
\bibliography{apssamp}
\bibliographystyle{apsrev4-2}

\end{document}